\documentclass[11pt]{article}
\usepackage{moriond,epsfig}
\usepackage{verbatim} 

\newcommand{\be}{\begin{equation}}
\newcommand{\ee}{\end{equation}}
\newcommand{\bea}{\begin{eqnarray}}
\newcommand{\eea}{\end{eqnarray}}
\newcommand{\ben}{\begin{enumerate}}
\newcommand{\een}{\end{enumerate}}
\newcommand{\nn}{\nonumber}
\newcommand{\crn}{\nonumber \\}

\newcommand{\la}{\lambda}

\newcommand{\fr}{\frac}
\newcommand{\bc}{\begin{center}}
\newcommand{\ec}{\end{center}}

\newcommand{\ggbbH}{gg\to b\bar{b}H}

\begin{document}
\vspace*{0.1cm}\rightline{LAPTH-Conf-1248/08}
\vspace*{4cm}
\title{Yukawa corrections to Higgs production associated with two bottom quarks at the LHC}
\author{ Le Duc Ninh }
\address{LAPTH, Universit\'e de Savoie, CNRS; BP 110, F-74941 Annecy-le-Vieux Cedex, France}
\maketitle\abstracts{
We investigate the leading one-loop Yukawa corrections to
the process $pp\rightarrow b\bar{b}H$ in the Standard Model. 
We find that the
next-to-leading order correction to the cross section is small about $-4\%$ if the Higgs mass is $120$GeV. 
However, the appearance of leading Landau singularity when $M_H\ge 2M_W$ can lead to a large 
correction at the next-to-next-to-leading order level for a Higgs mass around $160$GeV. 
}
\section{Introduction} 
The cross section for $b\bar{b}H$ production at the LHC is 
very small compared to the gluon fusion channel. However, it is important to study that because of 
the following reasons:
\begin{itemize}
\item It can provide a direct measurement of the bottom-Higgs Yukawa coupling ($\la_{bbH}$) which can be strongly 
enhanced in the MSSM.
\item We can identify the final state in experiment by tagging b-jets with high $p_T$. 
This reduces greatly the QCD background. 
\item Theoretically, it is a $2\to 3$ process at the LHC which is a good example of one-loop multi-leg calculations. Moreover, the process 
$\ggbbH$ is, to the best of our knowledge, the most beautiful example where the leading Landau singularity (LLS) occurs in an electroweak box Feynman diagram. Considering that one rarely encounters such a singularity, studying its effect is very important.  
\end{itemize}
The next-to-leading order (NLO) QCD correction to
the exclusive process $pp\to b\bar{b}H$ with high $p_T$ bottom quarks has been calculated
by two groups \cite{dittmaier_bbH}. The QCD correction is about $-22\%$ for $M_H=120$GeV and $\mu=M_Z$ 
(renormalisation/factorisation scale). No leading Landau singularity occurs in any QCD one-loop diagrams.

The aim of our
work is to calculate the Yukawa corrections, which are the leading electroweak
corrections in this case, to the exclusive $bbH$ final state with high $p_T$ bottom quarks at the LHC \cite{fawzi_bbH}. 
These corrections are triggered
by top-charged Goldstone loops whereby, in effect, an external $b$
quark turns into a top quark. Such type of
transitions can even trigger $g g \to b \bar b H$ even with
vanishing $\lambda_{bbH}$, in which
case the process is generated solely at one-loop level.
\section{Calculation and results} 
\begin{figure}[h]
\begin{center}
\includegraphics[width=12cm]{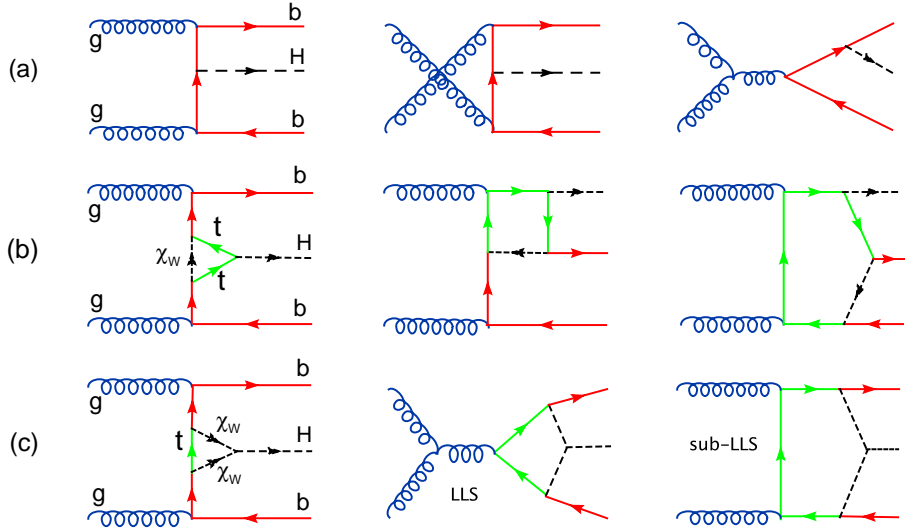} 
\caption{\label{diagrams}{\em Typical Feynman diagrams for the process $gg\to b\bar{b}H$ at tree [class (a)] and one-loop [classes (b) and (c)] levels. 
Loop particles are the charged Goldstone bosons ($\chi_W$) and the top quark. In the class (c): the box diagram has a LLS, the pentagon diagram has the sub-leading Landau singularity which is the same as the LLS of the box diagram. The LLS occurs when $M_H\ge 2M_W$ and $\sqrt{s}\ge 2m_t$, {\it i.e.} all the four particles in the loop can be simultaneously on-shell.}}
\end{center}
\end{figure}

\begin{figure}[t]
\begin{center}
\begin{minipage}[c]{6.5cm}
\includegraphics[width=6.5cm,height=4.5cm]{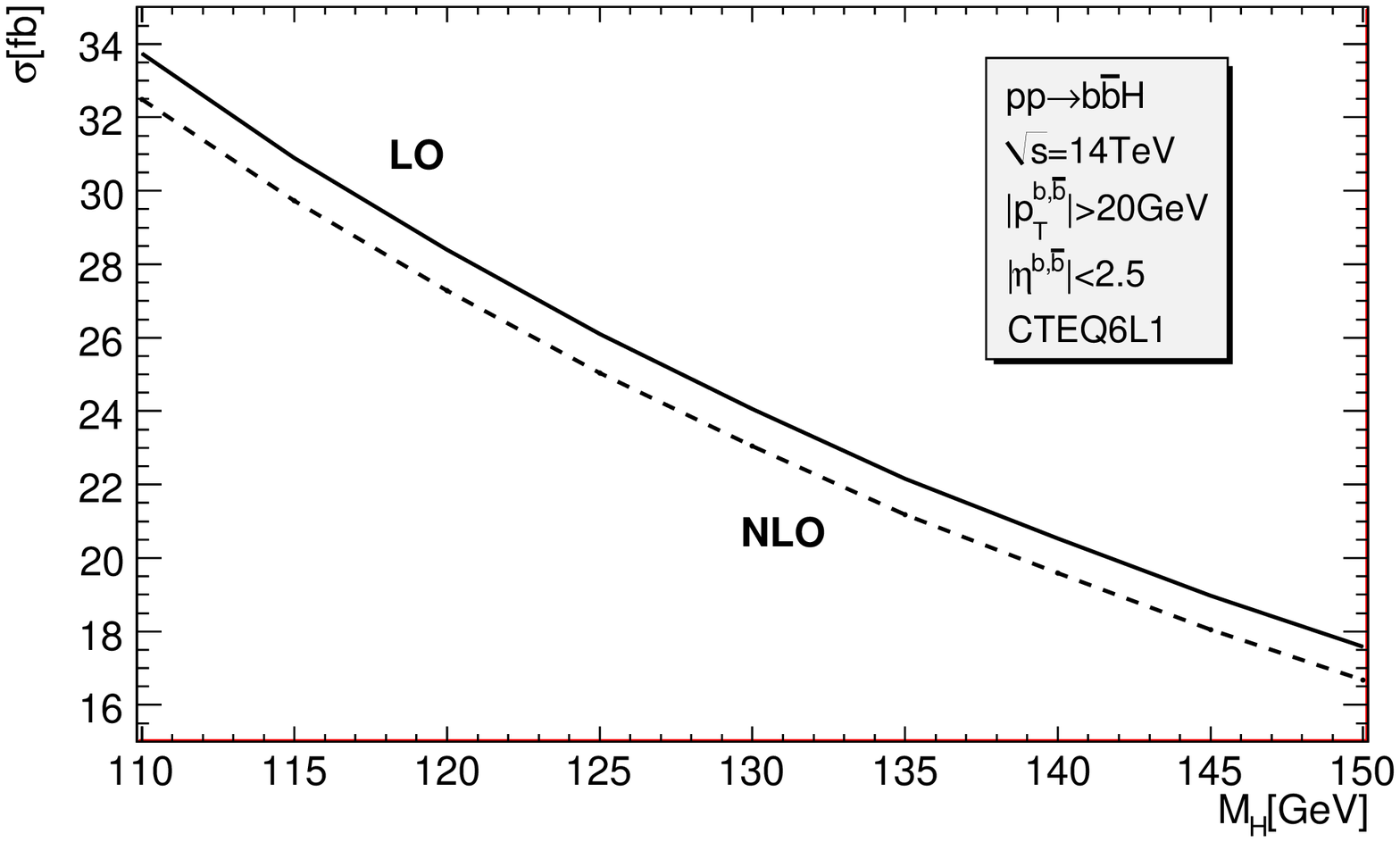}
\end{minipage}
\begin{minipage}[c]{6.5cm}
\includegraphics[width=6.5cm,height=4.5cm]{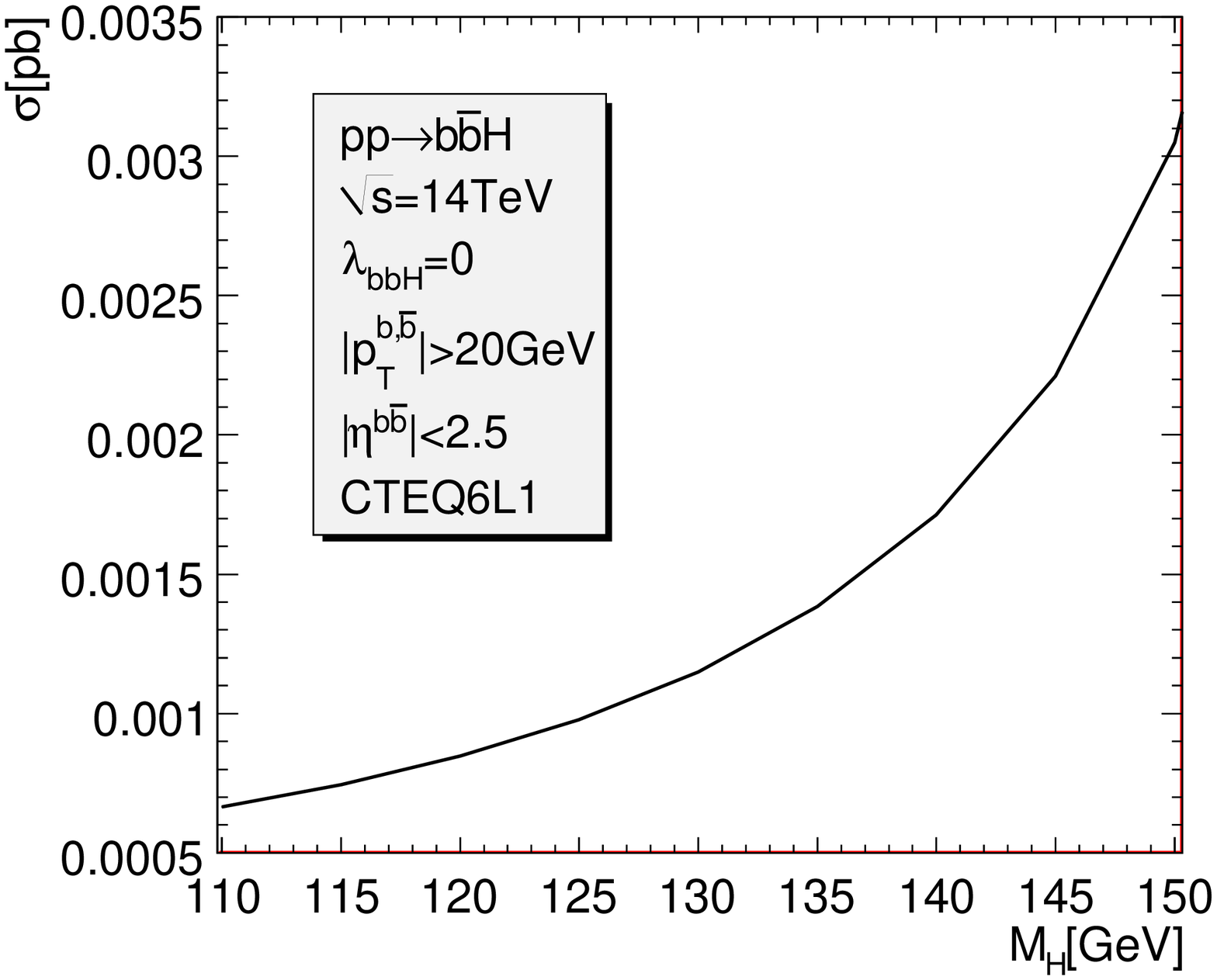}
\end{minipage}
\caption{\label{sigma_mH}{\em Left: the leading order (LO) and NLO cross sections
as functions of $M_H$. Right: the cross section in the limit of 
vanishing $\la_{bbH}$.}}
\end{center}
\end{figure}

At the LHC, the entirely dominant contribution comes 
from the sub-process $g g\to b \bar b H$. The contribution from the light 
quarks in the initial state is therefore neglected in our calculation. 
Typical Feynman diagrams at the tree and one-loop levels are 
shown in Fig. \ref{diagrams}. 
All the relevant couplings are:
\bea
\la_{bbH}&=&-\fr{m_b}{\upsilon},\hspace*{3mm} \la_{ttH}=-\fr{m_t}{\upsilon},\crn 
\la_{tb\chi}&=&-i\sqrt{2}\la_{ttH}(P_L-\fr{m_b}{m_t}P_R),\hspace*{3mm} 
\la_{\chi^+\chi^-H}=\fr{M_H^2}{\upsilon},\nn
\eea
where $\upsilon$ is the vacuum expectation value and $P_{L,R}=(1\mp\gamma_5)/2$. 
The cross section as a function of $\la_{bbH}$ can be written in the form
\bea \sigma(\la_{bbH})&=&\sigma(\la_{bbH}=0)+\la_{bbH}^2\sigma^\prime(\la_{bbH}=0)+\cdots,\crn
\la_{bbH}^2\sigma^\prime(\la_{bbH}=0)&\approx&\sigma_{NLO}=\sigma_{LO}[1+\delta_{NLO}(m_t,M_H)],\nn
\eea
where $\sigma(\la_{bbH}=0)$ is shown in Fig. \ref{sigma_mH} (right), $\sigma_{LO}$ and $\sigma_{NLO}$ are 
shown in the same figure on the left.  

\begin{figure}[h]
\begin{center}
\begin{minipage}[c]{6.5cm}
\includegraphics[width=6.5cm]{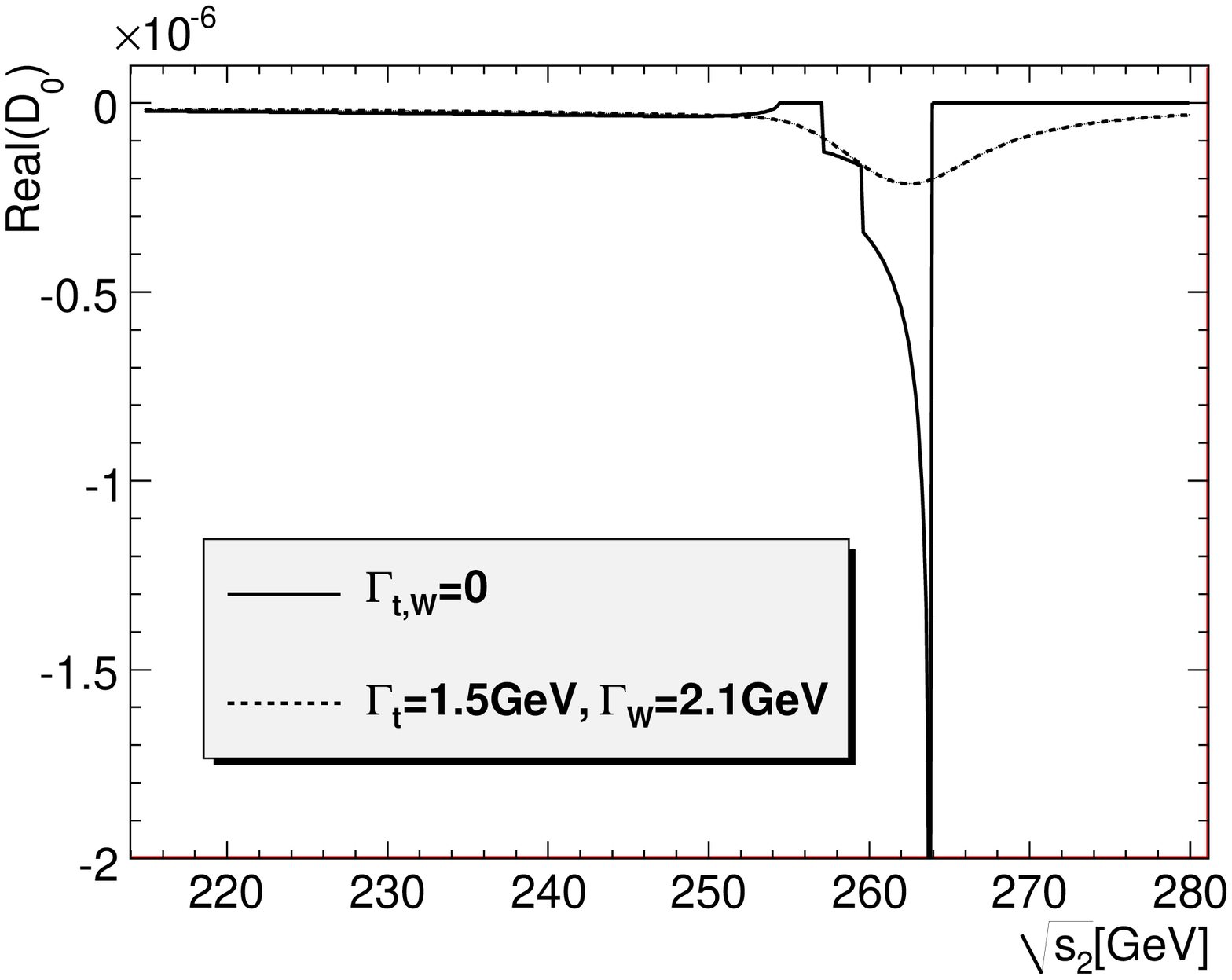}
\end{minipage}
\hspace*{0.1cm}
\begin{minipage}[c]{6.5cm}
\includegraphics[width=6.5cm]{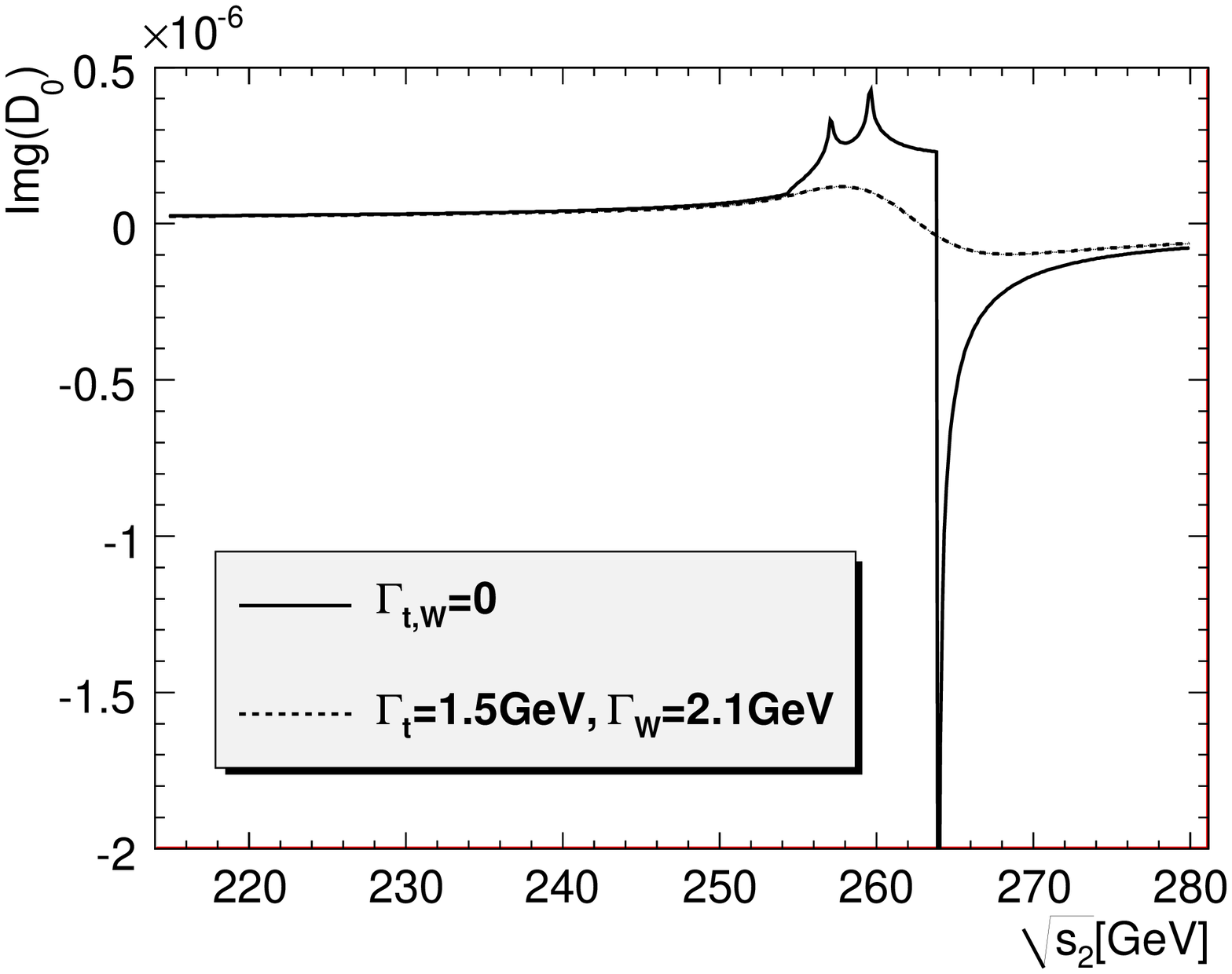}
\end{minipage}
\caption{\label{landau_sing}{\em The real and imaginary parts of the scalar box integral associated with the LLS diagram in the class (c) of Fig. \ref{diagrams}.}}
\end{center}
\end{figure}

$\sigma(\la_{bbH}=0)$ is generated solely at one-loop level and gets large when 
$M_H$ is close to $2M_W$. This is due to the 
leading Landau singularity related to the scalar loop integral associated to the box diagram in the class (c) of Fig. \ref{diagrams}.
This divergence, which occurs when $M_H\ge 2M_W$, is not integrable at the level of loop amplitude 
squared and must be regulated by introducing a width for the unstable particles in the loops. Mathematically, the width effect is to move 
the LLS into the complex plane so that they do not occur in the physical region. 
The solution is shown in 
Fig. \ref{landau_sing}. The important point here is that the LLS, even after being regulated, can 
lead to a large correction to the cross section, up to $49\%$ for $M_H=163$GeV, $\Gamma_W=2.1$GeV and $\Gamma_t=1.5$GeV.
%%%
\section*{Acknowledgments}
The author would like to acknowledge the financial support from EU Marie Curie Programme. This work has been done in collaboration with 
Fawzi Boudjema. 

\end{document}